DTU Library

# Equitable Transit Network Design Under Uncertainty

**Jiang, Yu**

# Equitable Transit Network Design Under Uncertainty

**Yu Jiang**

**Abstract** This paper proposes a bi-level transit network design problem considering supply-side uncertainty. The upper-level problem determines frequency settings to simultaneously maximise the efficiency and equity measures, which are defined by the reduction in the total effective travel cost and the minimum reduction in the effective travel cost of all OD pairs, respectively. The lower level problem is the reliability-based transit assignment problem that captures the effects of supply-side uncertainty on passengers' route choice behaviour. Numerical studies demonstrate that 1) the Pareto frontier may not be convex; 2) it is possible to improve the efficiency and equity objectives simultaneously; 3) increasing the frequency could worsen the equity measure; 4) passengers' risk attitude affects the rate of substitution between the two objectives.

## 1 Introduction

Equity is of major concern in designing of public transport systems (Delbosc and Currie, 2011) and it is of great importance to both travellers and planners. Because if passengers are treated unfairly, they would perceive the public transport system less attractive and, as a result, could switch their travel mode from public transport to private transport, amplifying the urban traffic congestion problems.

Equity, originally, is a concept in social science (Sen, 1973). Although there is no universally accepted definition, in the field of transport planning, it can generally be classified into horizontal and vertical equities (Xu et al., 2016). The horizontal

Yu Jiang
DTU Transport, Department of Management Engineering
Lyngby, Denmark
Email: yujiang@dtu.dk

equality focusses on efficiently move a large number of people, while the vertical equality concerns individual's accessibility needs. Both horizontal and vertical equities can be integrated into the transportation network design problem and various models have been developed to approach equitable designs of road network (Meng and Yang, 2002; Yang and Zhang, 2002; Chen and Yang, 2004; Szeto and Lo, 2006; Mollanejad and Zhang, 2014; Szeto et al., 2015, etc.)

In contrast, in the context of public transport, most existing studies focus on examining and evaluating the equity condition of a given transit network (Delbosc and Currie, 2011; Welch and Mishra, 2013; Foth et al., 2013; Kaplan et al. 2014; Wei et al., 2017, etc.), while only a few studies incorporate equity in the transit network design problems. Ferguson et al. (2012) developed an approach to provide equitable access to basic amenities via designing transit frequency. The genetic algorithm was adopted to solve the model. Camporeale et al. (2016) devised a multimodal network design problem, where the equity ratio established in Meng and Yang (2002) was generalised by taking into account the demand effect. Ruiz et al. (2017) proposed bus frequency optimisation methodology to improve social equity, which is measured by the Gini Index. Nevertheless, there are two critical issues that have not been addressed in the preceding literature. One is to capture passengers' response to the changes in the transit network via incorporating transit assignment model. Although Camporeale et al. (2016) encapsulated equilibrium constraints, the unique feature of the transit assignment model problem, i.e., the common line problem, was not considered. The other is to capture the effect of travel time uncertainty. Due to various factors such as road incidents, signal breakdown, and weather conditions, etc., the travel time components are indeed uncertain. The stochastic travel time will affect both the realisation of transit network design and passengers' route choice behaviour, resulting in different values of the equality measures.

To fill the above reach niches, this study will develop a bi-level framework to design equable transit network while considering the effect of supply-side uncertainty. The upper-level problem is the transit frequency design problem, while the lower level problem is the transit assignment problem. In the upper level, we adopt the spatial equality metric, which is defined by the ratio between the effective travel cost before and after the changes in the transit services, as the equity measure for an OD pair. Following Rawlsian principle stating that a justice society maximises the welfare of its worst-off members (Feldman and Kirman, 1974; Karsu and Morton, 2015), the equity objective is formulated to maximise the minimum improvement in the spatial equality metrics of all the OD pairs. In the lower level, the reliability-based transit assignment model developed in Jiang and Szeto (2016) is adopted. Their model captures passengers risk-aversion attitude over travel time uncertainty and can be solved efficiently by the extragradient method.

## 2 Formulation

We consider a general transit network, which is further transformed into the route-section network representation (de Cea and Fernández, 1993). Following the literature, the following assumptions are made. A1) Passengers arrive randomly, consider a set of attractive lines, and board the first arriving vehicle; A2) Stochastic vehicle headways with the same distribution function (i.e., exponential distribution) are assumed for vehicles serving different lines; A3) The travel demand between each OD pair in the system is assumed to be known; A4) The passenger selects the route that minimizes his/her effective travel cost, where the effective travel time composes the expected travel time and the safety margin which computed based on the variance of the travel time and passengers' risk-aversion attitude. Based on the proceeding assumptions, the transit network design problem is formulated as follows

$$\left[\overbrace{\max_{\mathbf{f}} \sum_{(r,d)\in Q}\left(\tilde{u}^{rd}-u^{rd}(\mathbf{f})\right)g^{rd}}^{\text{Efficiency Objective}},\overbrace{\max_{\mathbf{f}} \min_{(r,d)\in Q}\left\{\tilde{u}^{rd}-u^{rd}(\mathbf{f})\right\}}^{\text{Equity Objective}}\right]^{T} \tag{1}$$

Subject to

$$\sum_{l\in L} 2E\left[F^{l}\right]f^{l} \le V_{\max}, \forall l \in L \tag{2}$$

$$f_{\min} \le f^{l} \le f_{\max}, \forall l \in L \tag{3}$$

where $u^{rd}(\mathbf{f})$ is defined by

$$\left(\boldsymbol{\alpha}-\boldsymbol{\alpha}^{*}\right)\left(\mathbf{u}(\mathbf{f},\boldsymbol{\alpha})\right) \ge \mathbf{0}, \forall \boldsymbol{\alpha} \in \Omega \tag{4}$$

The upper-level problem, Eqs. (1) - (3), determines the frequency settings, while subjecting to the fleet size and the frequency boundary constraints. Equation (1) contains two objectives. The first one is the efficiency objective, which is to maximise the reduction in total passengers' effective travel cost, where $g^{rd}$ denotes the travel demand between nodes *r* and *d*, $\tilde{u}^{rd}$ is the equilibrium effective travel cost before the frequency settings are changed, and $u^{rd}(\mathbf{f})$ represents the equilibrium travel cost under frequency setting $\mathbf{f} = \left[f_l\right]$. The second objective is the equity objective, which is to maximise the minimum improvement in the equilibrium effective travel cost of all OD pairs. Constraint (2) is the fleet size constraint, where $E\left[F^{l}\right]$ is expected trip time associated with line *l*. Following Li et al. (2008) and Szeto et al. (2013), the expected trip time considers layover time, dwell time, travel time, and variance of travel time. Constraint (3) restricts the upper and lower bounds of the frequency variables.

The lower level problem, Eq.(4), is the variational inequality formulation for the reliability-based transit assignment problem using the concept of approach proportion. An approach of a node is defined by the route section coming out from that node, and an approach proportion is defined as the proportion of passengers leaving a node via the approach considered. $\boldsymbol{\alpha} = \left[\alpha_s^d\right]$ denotes the approach proportion and $\Omega$ is the solution space of the approach proportion. $\mathbf{u}(\mathbf{f}, \boldsymbol{\alpha}) = \left[u_s^{id}\right]$ is the mapping function from the approach proportion to the corresponding effective travel cost. The effective travel time, $u_s^{id}$ , is given by

$$u_s^{id} = E\left[C_s + C^{h(s)d}\right] + \rho Var\left[C_s + C^{h(s)d}\right], \forall s \in S, i \in N, d \in D \quad (5)$$

where $C_s$ is the travel cost associated with section *s*, $C^{h(i)d}$ is the minimum travel cost between the head node of section *s* and destination *d*, and $\rho$ represent the degree of passenger's risk aversion. By using the approach-based formulation, the transit assignment problem can be effectively solved by the extragradient method that only requires mild assumptions for convergence.

## 3 Numerical Examples

To investigate the properties the problem, the four-line network in de Cea and Fernández (1993), shown in Figure 1, was adopted as shown in Figure 1. Two OD pairs are considered, A-B and X-B. Without further specified, the parameters are set as: $g^{\text{AB}} = 175$ , $g^{\text{XB}} = 150$ , $\rho = 0.15$ , $f_{\min} = 4$ buses/hour, $f_{\max} = 15$ buses/hour, and $V_{\max} = 12$ buses.

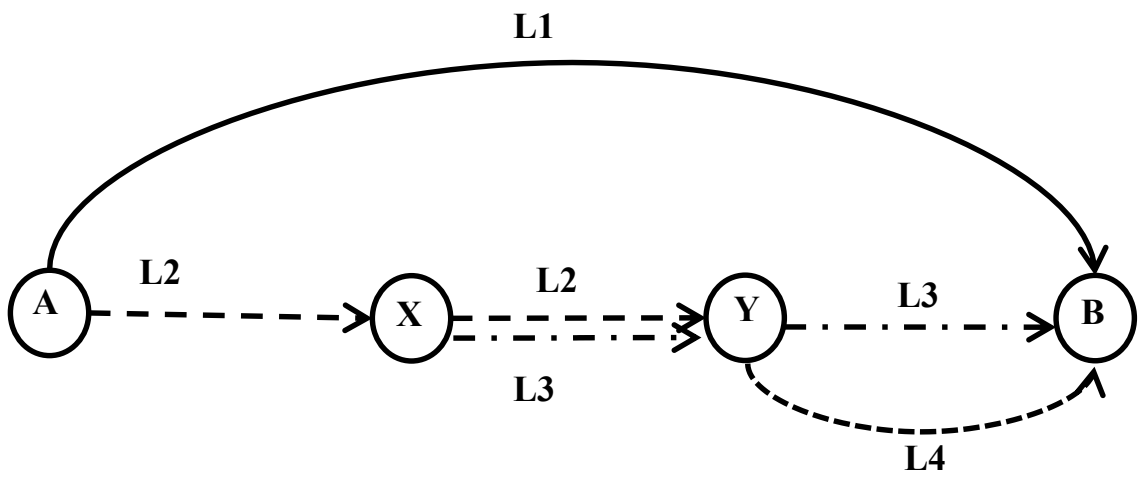

**Fig**. 1 Network

In the first experiment, the frequency of line L2 was varied from 4.0 to 9.0 buses per hour. Figure 2 plots the changes in the equity and efficiency objectives with respect to the changes in the frequency. It is observed that 1) the equity objective only improves within certain ranges of frequency settings, while the efficiency objective generally improves; 2) the two objectives could increase (decrease) simultaneously; the two objectives do not monotonically grow with the frequency.

The Pareto frontier is illustrated in Figure 3. Clearly, it shows that the Pareto frontier is not convex. Meanwhile, there is large gap between the rightmost point and its left neighbour, indicating that changing the equity objective between the two points has a significant effect on the change in the efficiency objective.

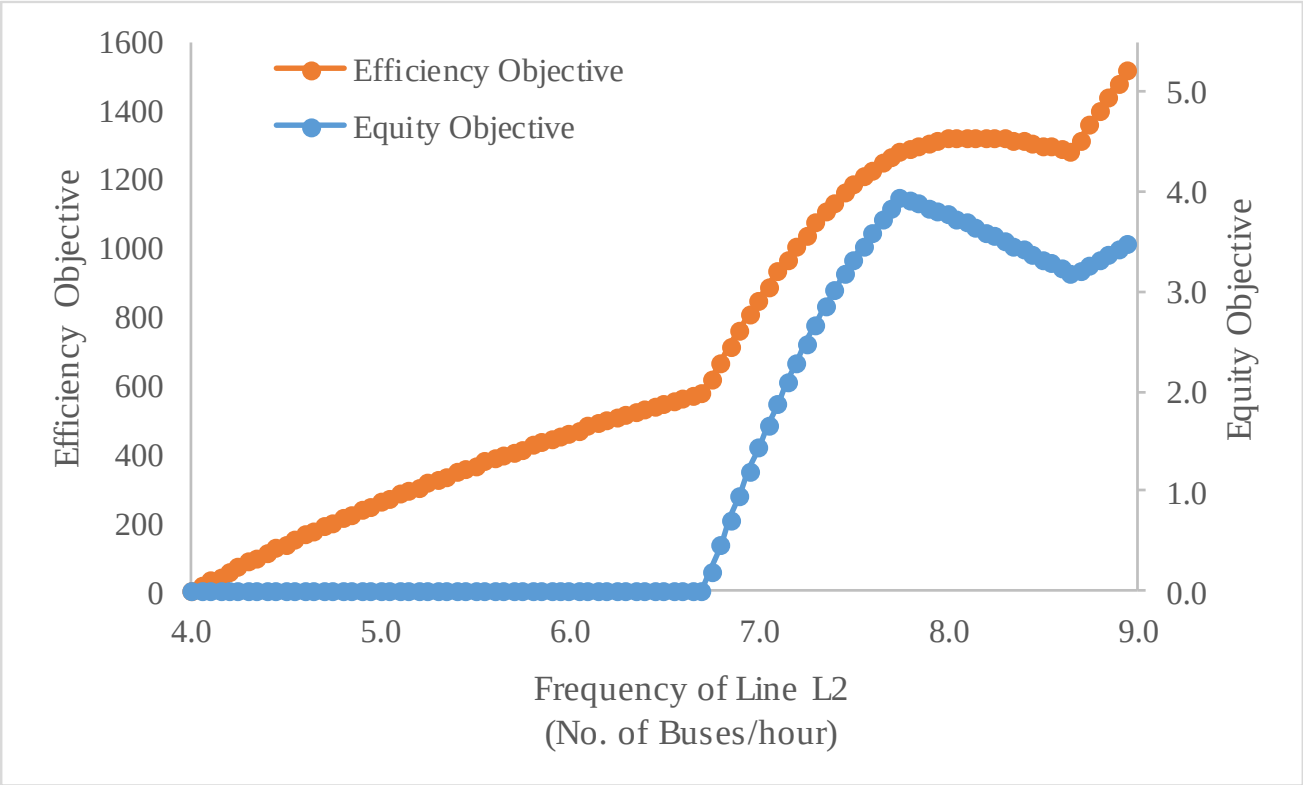

**Fig**. 2 Trade-off between the two objectives

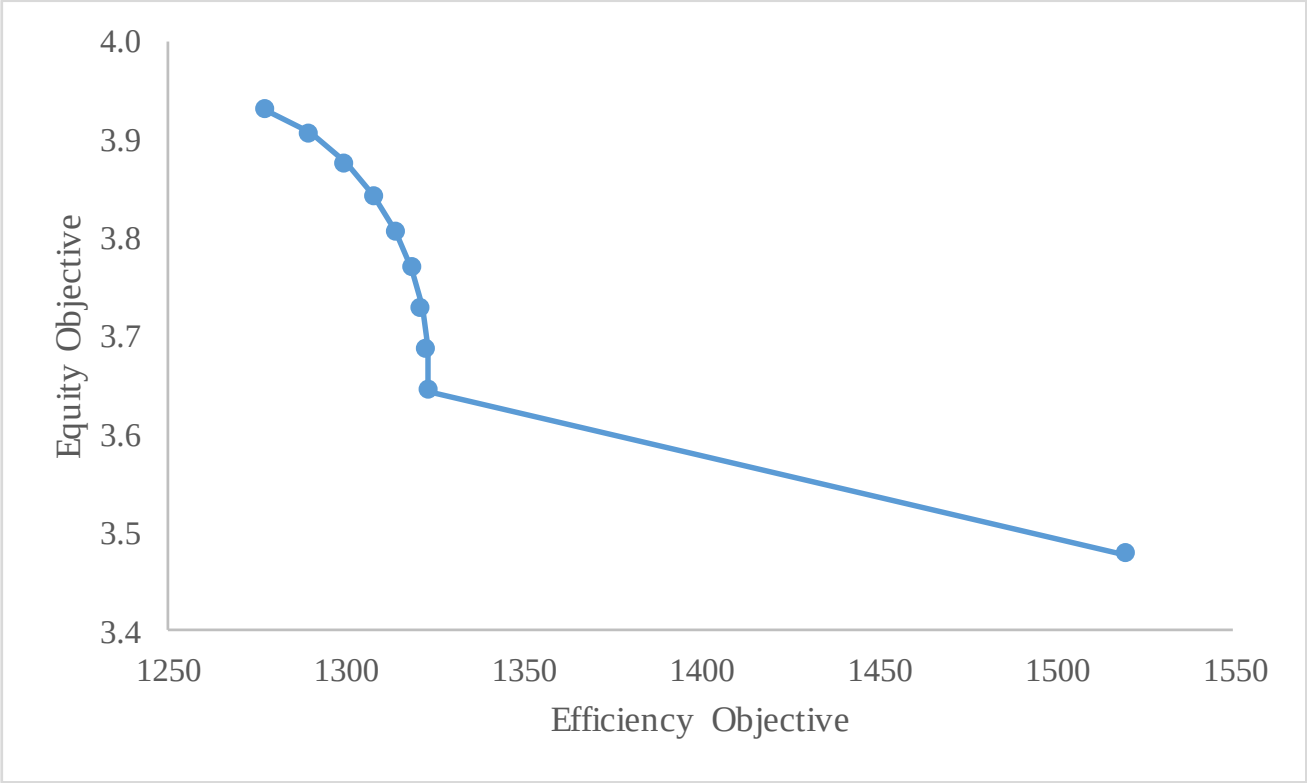

**Fig**. 3 Pareto frontier

To examine the effect of passengers' risk attitude on the Pareto frontier, $\rho$ was increased from 0.05 to 0.35 and the normalized[1] objective values are plotted in Fig. 4. Other than observing that the Pareto frontier could be convex (i.e., when $\rho = 0.05$ ), it is found that the degree of passenger's risk aversion affects the substitution rate between the efficiency and equity objectives.

[1] The normalized value is obtained by $x' = \left(x - \min(\mathbf{x})\right) \big/ \left(\max(\mathbf{x}) - \min(\mathbf{x})\right)$, where $x'$ is the normalized value and $x$ is the original value

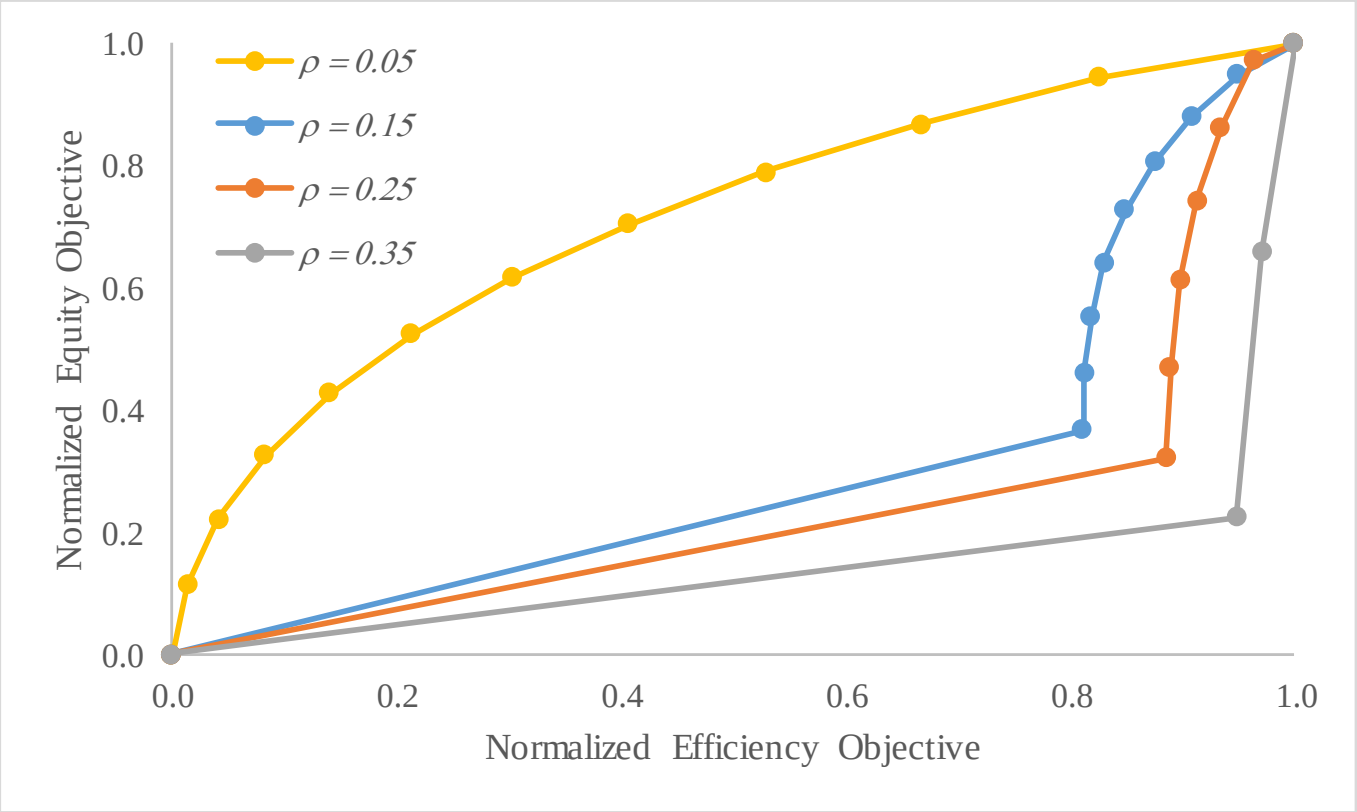

**Fig**. 4 Effect of $\rho$ on the Pareto frontier

The effect of the maximum fleet size on the Pareto frontier is illustrated in Fig. 5. It shows that the Pareto frontier moves towards the right-hand side and reduces to one point when $V_{\max} = 15$, indicating that the efficiency and equity objectives could be improved simultaneously. Nevertheless, it is also noticed that the maximum equity objectives are identical when $V_{\max}$ = 13, 14, and 15. The is due to the maximum frequency constraint, which restricts the maximum reduction in the effective travel cost between OD pairs.

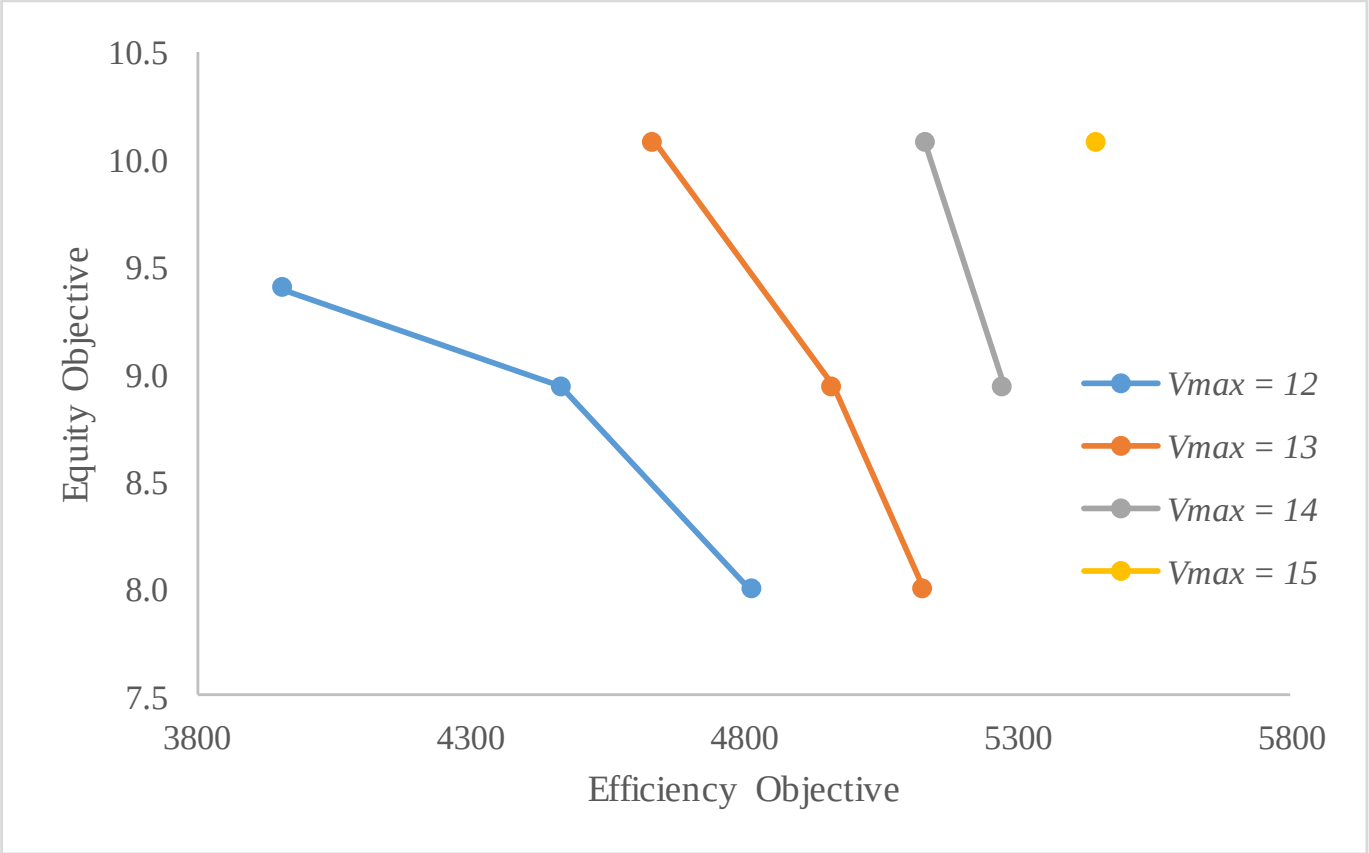

**Fig**. 5 Effect of $V_{\max}$ on the Pareto frontier

The effect of the maximum and minimum frequency on the Pareto frontier are illustrated in Figs. 6 and 7. Fig. 6 demonstrates that increasing the minimum frequency may not improve the two objectives. This is because a higher minimum frequency constraint could result in allocating fleet to the lines that are not efficient in term of reducing total passengers' effective travel cost. Fig. 7 shows that the larger the maximum allowable frequency, the higher values the two objectives achieve. Meanwhile, there is no trade-off when $f_{\max} = 13$ and $f_{\max} = 17$.

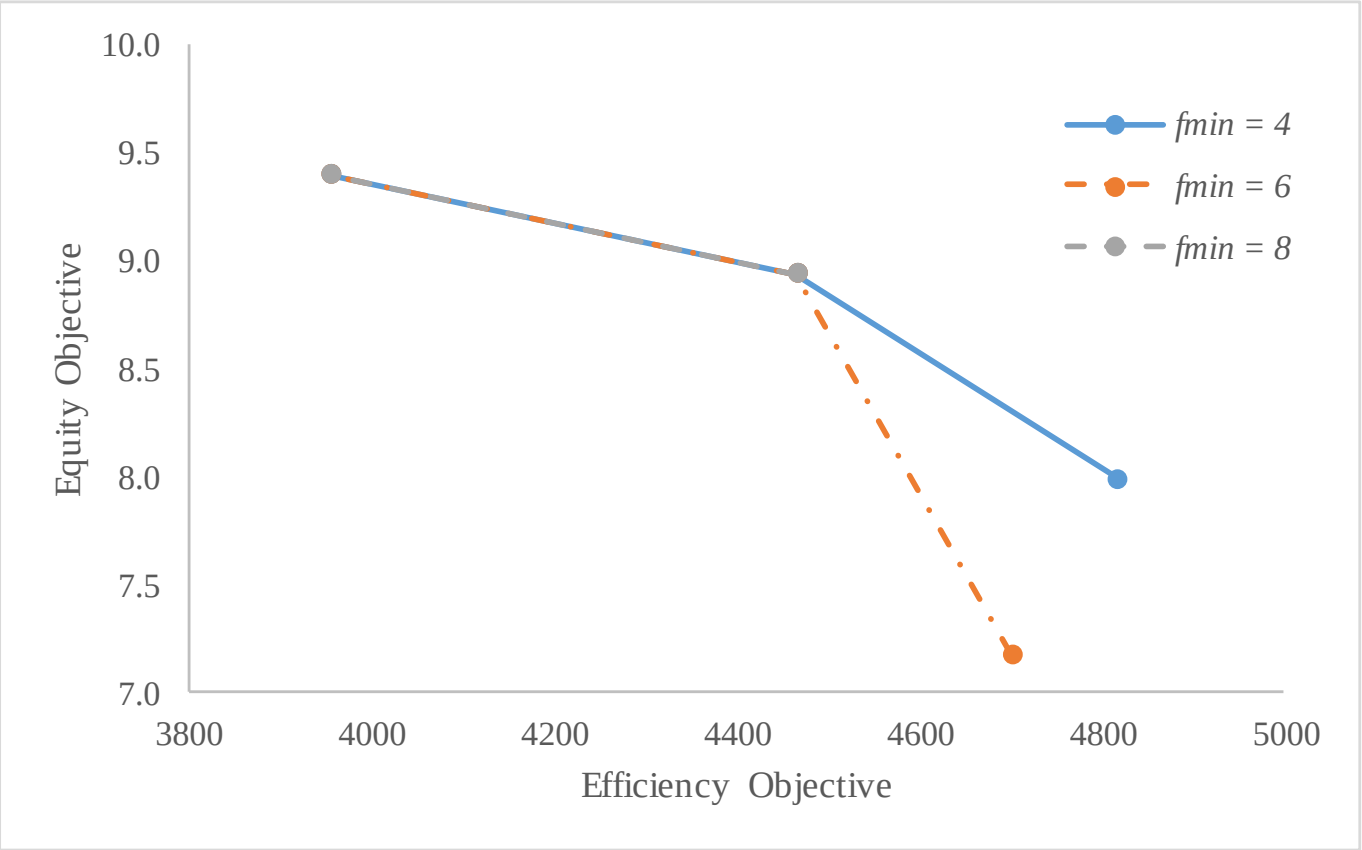

**Fig**. 6 Effect of $f_{\min}$ on the Pareto frontier

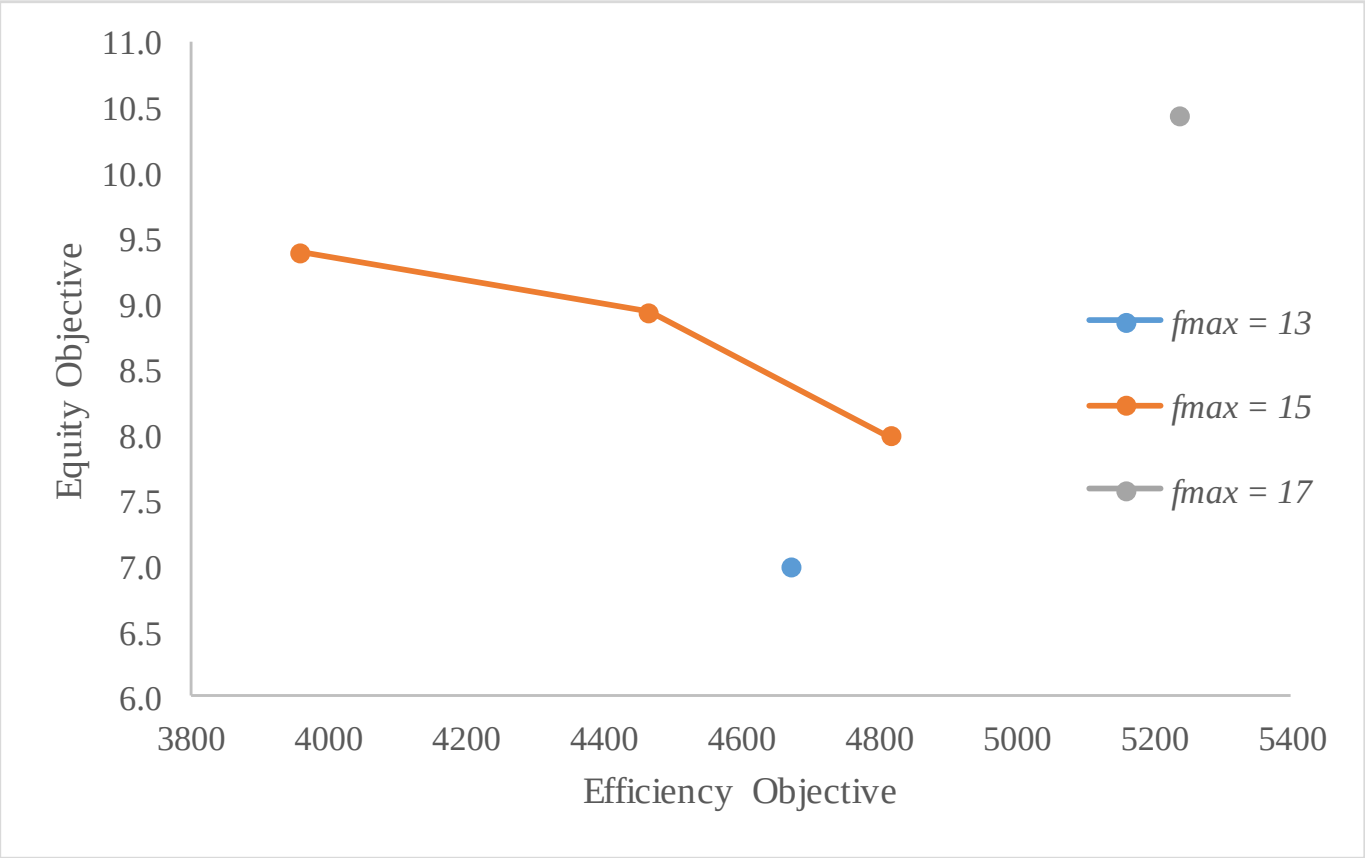

**Fig**. 7 Effect of $f_{\max}$ on the Pareto frontier

## 4 Conclusions

This paper developed a multi-objective bilevel formulation for the transit network design problem, in which equity is explicitly considered as an objective function, which is to maximise the minimum improvement in the effective travel cost among all OD pairs. The stochasticity of travel time on passengers' route choice behaviour is captured via the reliability-based transit assignment model which is formulated using the concept of approach proportion. The preliminary results illustrate the trade-off between the efficiency and equity objectives, the Pareto frontier may not be convex, and the two objectives could improve or decrease simultaneously. Future work will focus on developing an artificial bee colony algorithm (Szeto and Jiang, 2014) to solve the bilevel model.

**Acknowledgements:** This work is partly funded by the Innovation Fund Denmark (IFD) under File No. 4109-00005 and supported by National Science Foundation of China under Project No. 71701030.